\documentstyle[twoside,fleqn,espcrc2]{article}

\def\Haux{{\cal H}_{\rm aux}}
\def\Hphys{{\cal H}_{\rm phys}}
\def\Hphysprime{{{\cal H}'}_{\rm phys}}
\def\Aobs{{\cal A}_{\rm obs}}
\def\Phiphys{{\Phi_{\rm phys}}}
\def\Phiphysprime{{{{\Phi}'}_{\rm phys}}}
\def\L{{\cal L}}
\def\A{{\cal A}}                                                
\def\V{{\cal V}}
\def\inneraux{\langle\cdot\vert\cdot\rangle_{\rm aux}}
\def\innerphys{\langle\cdot\vert\cdot\rangle_{\rm phys}}

\title{Group Averaging and Refined Algebraic Quantization}

\author{D. Giulini\address{Institute for Theoretical Physics,
           University of Z\"urich,\\
           Winterthurerstrasse 190, CH-8057 Z\"urich, Switzerland}}%

\begin{document}

\begin{abstract}
We review the framework of Refined Algebraic Quantization
and the method of Group Averaging for quantizing systems
with first-class constraints. Aspects and results concerning 
the generality, limitations,  and uniqueness of these methods 
are discussed.
\end{abstract}

\maketitle

\section{Introduction}
Refined Algebraic Quantization (RAQ) is an attempt (amongst others) 
to concretize Dirac's program for the quantization of constrained 
systems within a generally applicable, well defined mathematical
framework.
It was first formulated as a general scheme in \cite{AAetal,Don1}
and recently developed further in \cite{DD1,DD2}. Here I wish to
report on these recent developments. 

The method itself has already been used (partly implicitly) earlier
in some successful applications to quantum gravity in specialized
situations, like linearized gravity on symmetric
backgrounds~\cite{Higuchi1,Higuchi2} or various minisuperspace
models~\cite{Don2,Don3}. These suggested the program to find a
general scheme of which these cases are just special cases.

Any scheme that deals with constrained systems needs to interpret
the phrase `solving the constraints'. Here the guiding idea of RAQ
is to work from the onset within an auxiliary Hilbert space, $\Haux$,
by means of which a $*$-algebra of observables, $\Aobs$, is constructed
{\it before} the constraints are `solved'. The $*$-operation on $\Aobs$
derives from the adjoint-operation $\dagger$ on $\Haux$, which allows to
connect the auxiliary inner product with the inner product on the
physical Hilbert space, $\Hphys$, since the latter is required to
support a $*$-representation of $\Aobs$. A possibly non-trivial
limitation derives from the fact that the constraint operators on
$\Haux$ are required to be self-adjoint, which may not be possible
in case they do not form a Lie-algebra (i.e. `close' with
structure-functions  only). This difficulty clearly does not arise
if the constraints derive from the action of a Lie-group $G$. In the 
following we shall restrict attention to such cases.  
{\it More precisely, we consider situations where a finite dimensional 
Lie group $G$ acts by some unitary representation $U$ on $\Haux$.}

\section{Concretizing the Dirac Procedure}
From the RAQ point of view, the concretization of the Dirac procedure
starts form an auxiliary Hilbert space $\Haux$ with inner product
$\inneraux$ and the set of unitary operators $\{U(g)\,g\in G\}$.
The naive reading of Dirac's prescription is to identify the space 
of physical states with those elements in $\Haux$ that are fixed by 
$G$:
\begin{equation}
U(g)\vert\psi\rangle=\vert\psi\rangle,\quad\forall g\in G,
\label{naive-dirac}
\end{equation}
which just says that one should pick the trivial subrepresentation 
of $U$. But clearly this statement needs not be well defined since 
$U$ might simply not contain the trivial representation as 
{\it sub}-representation. This will happen if the operators $U(g)$ do not 
all have the value `one' in the discrete part of their spectrum. 
Another difficulty comes in if $G$ is not unimodular.
Then it has been convincingly argued using methods of geometric 
quantization (see \cite{Duval-et-al} and references therein) 
that (\ref{naive-dirac}) is simply not the right condition, but 
that (\ref{naive-dirac}) should {\it formally} be 
replaced with 
\begin{equation}
U(g)\vert\psi\rangle=\Delta^{1/2}(g)\vert\psi\rangle\,,
\label{naive-dirac-nu}
\end{equation}
where $\Delta(g):={\rm det[Ad_g]}$ is the modular function on $G$ 
(a one-dimensional, real representation of $G$). We said `formally' 
since (\ref{naive-dirac-nu}) cannot hold in $\Haux$, for this 
would mean that the unitary operator $U(g)$ can change lengths by 
$\Delta^{1/2}(g)\not =1$, a plain contradiction. This is connected 
with the first problem since it means that $\Delta^{1/2}(g)$ 
cannot be a discrete spectral value of $U(g)$. In physical terms, 
$\vert\psi\rangle$ in (\ref{naive-dirac-nu}) is not normalizable. 
Hence one must read these equations in the appropriate sense.  
It is useful to break up the further development into several steps.

\medskip
\noindent
{\bf Step 1:}
We denote by $\dagger$ the adjoint map on operators on $\Haux$ 
with respect to $\inneraux$. Choose a dense linear subspace 
$\Phi\subseteq\Haux$ which is left invariant (as set) by $G$'s 
action. This choice of $\Phi$ is an important step and will 
generally require some physical input. Let $\Phi^*$ be the 
algebraic dual (linear functionals) of $\Phi$. Put on $\Phi^*$ 
the topology of pointwise convergence: 
$f_n\rightarrow f$ in $\Phi^*$ iff $f_n[\phi]\rightarrow f[\phi]$ 
as real numbers for all $\phi\in\Phi$. Note that, conversely, each 
$\phi\in\Phi$ defines a {\it continuous} linear functional on $\Phi^*$ 
by setting $\phi(f):=f[\phi]$. Hence $\Phi$ embeds in the topological
dual (continuous linear functionals) of $\Phi^*$.

\medskip
\noindent
{\bf Step 2:}
Consider the set $\L:=\{A\mid \Haux\supseteq D(A)\rightarrow\Haux\}$ of
linear operators, where $D(A)$ denotes the domain of $A$.
It is not an algebra due to mismatches of ranges
and domains. We define a subset of $\L$ by
\begin{eqnarray}
\A:=\{A\in\L\mid\Phi\subseteq D(A)\cap D(A^{\dagger})
\phantom{xxxx}
&&\nonumber\\
\Phi\supseteq A(\Phi),\ \Phi \supseteq A^{\dagger}(\Phi)\},
&&
\label{defA}
\end{eqnarray}
and make it into an algebra by restricting the action of each $A\in\A$
to $\Phi$. Since $\dagger$ also restricts to $\A$, we have a $*$-algebra
which -- without indicating the restriction to $\Phi$ -- we continue to
call $\A$. Note that, by definition of $\Phi$, $\A$ contains the operators 
$U(g)$. {\it From now on, the $*$-operation of this algebra is the
only trace left by $\inneraux$}. Finally, we define a sub-$*$-algebra
$\Aobs\subseteq\A$ as the `commutant' in $\A$ of the set 
$\{U(g)\ g\in G\}$:
\begin{equation}
\Aobs:=\{A\in\A\mid U(g)A=AU(g),\ \forall g\in G\}.
\label{defAobs}
\end{equation}

\medskip
\noindent
{\bf Step 3:} 
Each $A\in\A$ acts as continuous linear map on $\Phi^*$ via the 
`adjoint' action: 
\begin{equation}
Af:=f\circ A^* .
\label{defadjointaction}
\end{equation}
Hence the constraint operators also act on $\Phi^*$ and we can 
define the {\it solution set}, $\V$, by
\begin{equation}
\V:=\{f\in\Phi^*\mid U(g)f=\Delta^{1/2}(g)f\},
\label{defsolutionset}
\end{equation}
which, by construction, carries an anti-$*$-representation of 
$\Aobs$ by continuous maps. (The `anti' is a consequence of 
the $*$ in (\ref{defadjointaction}) and does no harm)

\medskip
\noindent
{\bf Step 4:}
The Hilbert space of physical states is now to be found within $\V$.
Hence we seek a subspace $\Hphys\subseteq\V$ with inner product 
$\innerphys$ that makes $\Hphys$ into a Hilbert space. Generally one 
cannot turn all of $\V$ into a Hilbert space, since this would imply 
that all $A\in \Aobs$ were defined everywhere on $\Hphys$ and hence 
bounded, which is too restrictive. 
(Note: Merely having the whole Hilbert space as domain does not
yet imply boundedness of a linear operator $A$. But if $A$ 
{\it and} its adjoint are defined everywhere boundedness follows.)
Hence $\Hphys$ will in general 
be a proper subset of $\V$ and the topology of $\Hphys$ induced by 
$\innerphys$ must be finer than that it inherits from $\Phi^*$, 
since the former must be closed. This also means that operators 
in $\Aobs$ will not necessarily be bounded on $\Hphys$ and hence 
we do not have an anti-$*$-representation of $\Aobs$ on $\Hphys$
since domains and ranges might not match. Hence we proceed 
as usual by assuming that there is a dense subspace 
$\Phiphys\subseteq\Hphys$ on which we have such a 
representation. But besides these technicalities the relevant 
condition on $\innerphys$ is this: {\it the physical inner 
product is to be chosen such that for all $A\in \Aobs$ we have 
$A^*=A^{\dagger}$ on $A^*$'s domain in $\Hphys$, where $\dagger$ 
now denotes the adjoint operation with respect to $\innerphys$. 
This is how $\innerphys$ is influenced (but generally not determined) 
by $\inneraux$. It is with respect to this adjoint operation that
we speak of an anti-$*$-representation of $\Aobs$ on $\Phiphys$.}

\section{RAQ and the $\eta$-Map}
RAQ concretizes step~4 of the Dirac procedure just outlined.
It aims to construct $\Hphys$ by finding a so-called $\eta$-map
(or `rigging map'), which is an antilinear map $\eta:\Phi
\rightarrow\Phi^*$ such that the image of $\eta$ consists 
entirely of solutions, i.e., $\eta(\Phi)\subseteq\V$. 
Further conditions on this map are
\medskip

1. $\eta$ is real: $\eta(\phi_1)[\phi_2]=
\overline{\eta(\phi_2)[\phi_1]}$

2. $\eta$ is positive: $\eta(\phi)[\phi]\geq 0$

3. $\eta$'s image is an invariant domain for $\Aobs$ and $\eta$ 
intertwines the representations of $\Aobs$ on $\Phi$ and 
$\Phi^*$:  
\begin{equation}
A\eta(\phi)=\eta(A\phi)
\label{eta-intertwines}
\end{equation}

Given such an $\eta$, one defines $\innerphys$ on its image by
\begin{equation}
\langle\eta(\phi_1)\vert\eta(\phi_2)\rangle_{\rm phys}:=
\eta(\phi_2)[\phi_1]
\label{definnerphys}
\end{equation}
and then defines $\Hphys$ as the completion of $\eta$'s image with 
respect to the uniform structure (Cauchy sequences) defined by this 
inner product. {\it By construction, $\innerphys$ satisfies the 
condition that the $*$-operation on $\Aobs$ is the adjoint with 
respect to $\innerphys$.} 

We need only check two points in order to see that the 
$\Hphys$ so defined satisfies the conditions of step~4: 
First, we wanted $\Hphys$ to be a subset of $\V$, so we need to 
check that the completion just mentioned does not add points outside
$\V$, i.e.,  that $\Hphys$ is indeed a subset of $\V$ with finer 
intrinsic topology. To see this, we consider the map 
$\sigma:\Hphys\rightarrow\Phi^*$, defined by 
\begin{equation}
\sigma(f)[\phi]:=\langle\eta(\phi)\vert f\rangle_{\rm phys}\,,
\label{defsigma}
\end{equation}
and note immediately that $\sigma(f)$ vanishes iff $f$ is orthogonal
to all elements in the images of $\eta$. But since the image is 
dense in $\Hphys$ by construction $f$ must itself vanish, hence 
$\sigma$ is injective. Next we prove that $\sigma$ is continuous: 
if $f_n\rightarrow f$ in $\Hphys$ then 
$\langle\eta(\phi)\vert f_n\rangle_{\rm phys}\rightarrow
\langle\eta(\phi)\vert f\rangle_{\rm phys}$ for all $\phi$, 
since $\langle\vert\eta(\phi)\vert\cdot\rangle_{\rm phys}$ 
is a continuous linear form on 
$\Hphys$. Hence $\sigma(f_n)[\phi]\rightarrow\sigma(f)[\phi]$ for 
all $\phi$ which, since $\Phi^*$ carries the topology of pointwise 
convergence, implies $\sigma(f_n)\rightarrow\sigma(f)$ and therefore 
continuity of $\sigma$. 

The second point to check is that we indeed have an 
anti-$*$-representation of $\Aobs$ on a dense subspace $\Phiphys$ 
in $\Hphys$. To show this, we simply identify $\Phiphys$ with the 
image of $\eta$ and note that by (\ref{defadjointaction}) and 
(\ref{eta-intertwines}) we have for all $\phi\in\Phi$:
\begin{eqnarray}
\sigma(Af)[\phi] 
&=& \langle\eta(\phi)\vert Af\rangle_{\rm phys}
    =\langle\eta(A^*\phi)\vert f\rangle_{\rm phys}\nonumber\\
&=& \sigma(f)[A^*\phi]=A\sigma(f)[\phi].
\label{sigmaintertwines}
\end{eqnarray}
Hence $\sigma:\Phiphys\rightarrow\hbox{Image}(\sigma)\subseteq\Phi^*$ is 
an isomorphism of anti-$*$-representations of $\Aobs$. 

An interesting question is how general this method of constructing 
$\Hphys$ via an $\eta$-map actually is; that is, given $\Hphys$ as
in step~4 above, is there always an $\eta$ map whose image is 
$\Hphys$? Well, remember that each $\phi\in\Phi$ defines a continuous
linear functional on $\Phi^*$. Restriction to the linear subspace 
$\Hphys$ of $\Phi^*$ yields a continuous linear functional on 
$\Hphys$ with respect to its intrinsic topology, since the
latter 
is finer than that induced by $\Phi^*$. Hence, by Riesz' theorem,
there is a a unique vector $\eta'(\phi)\in\Hphys$ which satisfies
\begin{equation}
\phi(f):=f[\phi]=\langle\eta'(\phi)\vert f\rangle_{\rm phys}
\label{Riesz}
\end{equation}
for each $f\in\Hphys$. The map $\eta':\Phi\rightarrow\Hphys$ is 
obviously antilinear and has a non-trivial image, since
$\hbox{kernel}(\eta')=\bigcap_{f\in\V}\hbox{kernel}(f)\not =\Phi$
for $\V\not =\{0\}$. 
We can now define $\Phiphysprime$ to be the image of this $\eta'$ and 
$\Hphysprime\subseteq\Hphys$ its completion. This {\it almost} proves 
that we can at least reproduce the subspace $\Hphysprime$ by an 
$\eta$-map if $\eta'$ were indeed an $\eta$-map in the technical 
sense. But it might fail condition~3 above  that it intertwines 
with $\Aobs$; the reason being that so far nothing ensures 
$\Phiphysprime$ to be contained in $\Phiphys$. Hence operators in 
$\Aobs$ might not act on $\Phiphysprime$ at al! However, if
$\Phiphysprime\subseteq\Phiphys$ then $\eta'$ is indeed an $\eta$-map 
and we can at least claim to be able to reconstruct {\it some} sector      
$\Hphysprime\subseteq\Hphys$. 

Given that, we may finally wonder what additional assumptions would 
guarantee that we can construct {\it all} of $\Hphys$ by RAQ. One such 
condition is the following \cite{DD1}: {\it If $\Hphys$ constructed by 
the Dirac procedure decomposes into a direct sum of superselection 
sectors, then each sector can be separately constructed by the 
Dirac procedure.} If this is satisfied we argue as follows: suppose 
$\Hphysprime$ is a maximal subsector of $\Hphys$ which can be 
constructed by RAQ. Then its orthogonal complement is also a 
sector which, by hypothesis, is separately constructible by the Dirac 
procedure. But then, again by hypothesis, we can reconstruct a 
subsector of this orthogonal complement, which contradicts the 
assumption that $\Hphysprime$ was maximal.

\section{Group Averaging}
Whereas the $\eta$-map is a specific way to build $\Hphys$ for 
step~4 in the Dirac procedure, Group Averaging is in turn a 
specific way to construct an $\eta$-map. 
In Dirac's $\langle\hbox{bra}\vert\hbox{ket}\rangle$-terminology, 
the idea is simply to define (and make sense of)
\begin{equation}
\eta\vert\phi\rangle:=\int_Gd\mu(g)\,\langle\phi\vert U(g)\,,
\label{defgroupaveraging}
\end{equation}
where $d\mu$ is some appropriately chosen  measure on $G$. 
If $G$ were unimodular, the 
right- and left-invariant Haar measures coincide and are the 
correct choice for $d\mu$ in order for $\eta$'s image to formally 
solve (\ref{naive-dirac}) {\it and} make $\eta$ real in the sense of 
condition~1 above. Note that this reality condition requires the 
measure $d\mu(g)$ to be invariant under the inversion map 
$I: g\rightarrow g^{-1}$. For non-unimodular groups neither the
left- nor the right-invariant measure is $I$-invariant. 
Instead one has to take the `symmetric' measure, defined by    
\begin{eqnarray}
d\mu_0(g):&=&\Delta^{1/2}(g)\, d\mu_L(g)\nonumber\\
          &=&\Delta^{-1/2}(g)\, d\mu_R(g)\,,
\label{defsymmetricmeasure}
\end{eqnarray}
where $d\mu_L,d\mu_R$ are the left- and right-invariant measures 
respectively. Moreover, this measure also implies that $\eta$'s
image (formally) satisfies the modified 
equation~(\ref{naive-dirac-nu}). 
We refer to \cite{DD2} for a more extensive discussion of the 
modified Dirac condition and how it can be derived by Group 
Averaging from an appropriate adaptation of the 
`unimodularization' technique originally developed in geometric 
quantization~\cite{Duval-et-al}.

The physical inner product according to Group Averaging is given by 
\begin{equation}
\langle\eta(\phi_2)\vert\eta(\phi_1)\rangle_{\rm phys}:=
\int_Gd\mu_0(g)\langle\phi_1\vert U(g)\phi_2\rangle_{\rm aux}
\label{ga-innerproduct}
\end{equation}
so that the Group Averaging procedure only makes sense for states 
$\phi_{1,2}$ for which this integral converges absolutely.  
But it turns out \cite{DD2} that a more restricted choice of $\Phi$ 
is appropriate: we say that $\phi$ is an $L^1$ state, iff 
\begin{equation}
\int_gd\mu_n\langle\phi\vert U(g)\phi\rangle_{\rm aux}
\label{n-ga-innerproduct}
\end{equation}
converges absolutely for all integers $n$, where 
$d\mu_n(g):=\Delta^{n/2}(g)d\mu_0$.
  One can then also construct the group algebra 
$\A_G$ of functions on $G$ which are $L^1$ with respect to 
$d\mu_n$ for all $n$, and prove that its action on $L^1$ 
states results again in $L^1$ states . This fact allows to 
prove a uniqueness theorem in the following form (see \cite{DD2} for 
more details):  

\newtheorem{theorem}{Theorem}
\begin{theorem}
Suppose $\Phi$ is an $L^1$ subspace of $\Haux$  which is 
invariant under $\A_G$. Then, up to overall scale, any 
$\eta$-map is of the form (\ref{defgroupaveraging}) 
with $d\mu=d\mu_0$.
\end{theorem}

\section{Summary}
We outlined two technical devices which, to a certain extent, 
concretize the Dirac approach to quantizing constrained systems 
along the sequence [Group Averaging] $\rightarrow$ [$\eta$-map] 
$\rightarrow$ [Dirac approach]. There are fairly strong 
uniqueness theorems regarding these devices, provided Group 
Averaging converges sufficiently rapidly. But there is 
no general characterization when this can actually be achieved. 
Physical as well as mathematical inputs are required, perhaps 
most importantly in the choice of $\Phi$, and results are likely 
to delicately depend on this choice. Recent investigations of 
convergence properties in contexts of specific models 
\cite{Gomberoff-Marolf,Louko-Rovelli} confirm this general 
expectation.

Finally we mention the obvious, namely that the restriction 
to systems whose first class constraints close with structure 
{\it constants} (i.e. form a Lie algebra) should be lifted.
After all, GR is not in this form. We plan to investigate this 
in the future.

\end{document}